# GRVI Phalanx: A Massively Parallel RISC-V FPGA Accelerator Accelerator

Jan Gray, Gray Research LLC
Bellevue, WA, USA
jsgray@acm.org

*Abstract*— GRVI is an FPGA-efficient RISC-V RV32I soft processor. Phalanx is a parallel processor and accelerator array framework. Groups of processors and accelerators form shared memory clusters. Clusters are interconnected with each other and with extreme bandwidth I/O and memory devices by a 300-bit-wide Hoplite NOC. An example Kintex UltraScale KU040 system has 400 RISC-V cores, peak throughput of 100,000 MIPS, peak shared memory bandwidth of 600 GB/s, NOC bisection bandwidth of 700 Gbps, and uses 13 W.

## I. INTRODUCTION

In this Autumn of Moore's Law, the computing industry is challenged to scale up throughput and reduce energy. This drives interest in FPGA accelerators, particularly in datacenter servers. For example, the Microsoft Catapult [1] system uses FPGA acceleration at datacenter scale to double throughput or cut latency of Bing query ranking.

As computers, FPGAs offer parallelism, specialization, and connectivity to modern interfaces including 10-100 Gbps Ethernet and many DRAM channels (soon HBM). Compared to general purpose CPUs, FPGA accelerators can achieve higher throughput, lower latency, and lower energy.

There are two big challenges to development of an FPGA accelerator. The first is software: it is expensive to move an application into hardware, and to maintain it as code changes. Rewriting C++ code in RTL is painful. High level synthesis maps a C function to gates, but does not help compose modules into a system, nor interface the system to the host.

OpenCL to FPGA tools are a leap ahead. Now developers have a true software platform that abstracts away low level FPGA concerns. But "OpenCL to FPGA" is no panacea. Much software is not coded in OpenCL; the accelerator is specialized to particular kernel(s); and a design spin may take hours.

To address the diversity of workloads, and for fast design turns, more of a workload might be run directly as software, on processors in the fabric. (Soft processors may also be tightly coupled to accelerators.) To *outperform* a full custom CPU requires *many* FPGA-*efficient* soft processors.

The second challenge is implementation of the accelerator SOC hardware. The SOC consists of dozens of compute and accelerator cores, interconnected to each other and to extreme bandwidth interface cores e.g. PCI Express, 100G Ethernet, and, in the coming HBM era, eight or more DRAM channels. How is it possible to interconnect the many compute and interface cores at full bandwidth (50+ Gbps)?

GRVI Phalanx is a framework for building accelerators that helps address these problems. A designer can use any mix of software, accelerated software, or fixed function accelerators, composed together into clusters. With a cluster, cores can be directly coupled or communicate through shared memory. Between clusters and external interface cores, a wide Hoplite NOC [2] carries data as messages at full bandwidth.

The rest of this **extended abstract** details a work-in-progress GRVI Phalanx system, bottom up, from the GRVI soft processor, to the cluster architecture, to the full system.

## II. THE GRVI RISC-V RV32I CORE

Actual *acceleration* of a software-mostly workload requires an FPGA-efficient soft processor that implements a standard instruction set architecture (ISA) for which the diversity of open source software tools, libraries, and applications exist. The RISC-V ISA [3] is a good choice. It is an open ISA; it is modern; extensible; designed for a spectrum of use cases; and it has a comprehensive infrastructure of open source specifications, test suites, compilers, tools, simulators, libraries, operating systems, and processor and interface IP. Its core ISA, RV32I, is a simple 32-bit integer RISC. There are over a dozen extant FPGA soft processor implementations. A question for the present work was: is it possible to devise an extremely FPGA-efficient implementation of RISC-V RV32I?

A processor design is an exercise in which features to include and which to leave out. A simple, austere core uses fewer resources, which enables more cores per die, achieving more compute and memory parallelism per die.

The design goal of the GRVI RV32I core was therefore to maximize MIPS/LUT. This is achieved by eliding inessential logic from each CPU. Infrequently used resources, such as shifter, multiplier, and byte/halfword load/store, are cut from the core. Instead, they are shared by two or more cores in the cluster, so that their overall amortized cost is at least halved.

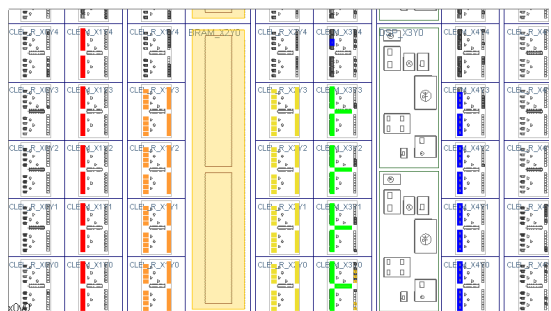

Fig. 1: GRVI processing element (PE) datapath RPM

The GRVI microarchitecture is unremarkable. It is a two or three stage pipeline (optional instruction fetch latch; decode; execute). The datapath (Fig. 1) includes a 2R/1W register file; two sets of operand multiplexers (operand selection and result forwarding) and registers; an ALU; a dedicated comparator for conditional branches and SLT (set less than); a PC unit for I-fetch, jumps, and branches; and a result multiplexer to select a result from ALU, return address, load data, optional shift and/or multiply. The color shaded LUTs in Fig. 1, left to right, comprise the register file, first and second operand muxes and registers. ALU, and result multiplexer – all 32-bits wide.

The datapath 6-LUTs are explicitly technology mapped and floorplanned into a relationally placed macro (RPM). The ALU, PC unit, and comparator use "carry logic". Each LUT in the synthesized control unit is scrutinized.

GRVI is small and fast. The datapath uses 250 LUTs; the core overall is typically 320 LUTs. It runs at up to 375 MHz in Kintex UltraScale (-2). Its CPI is not yet determined but is expected to be ~1.3 / ~1.6 (2 / 3 stage). Thus the figure of merit for the core is approximately 0.7 MIPS/LUT.

### III. GRVI CLUSTERS

As GRVI is relatively compact, it is possible to implement many processing elements (PEs) per FPGA – 750 in a 240,000 LUT Xilinx Kintex UltraScale KU040. Of course, PEs need memories and interconnects. A KU040 has 600 dual-ported 1Kx36 BRAMs. There are many ways to apportion them to PEs. In the present design, a cluster of eight PEs share 12 BRAMs. Four BRAMs are used as small 4 KB kernel program instruction memories (IRAMs); each pair of processors share one IRAM. The other eight BRAMs form a 32 KB cluster shared memory (CRAM). This has 12 32-bit wide ports. Four ports provide a 4-way banked interleaved memory for PEs. Each cycle, up to four accesses may be made on the four ports. The remaining eight ports provide an 8-way banked interleaved memory to an optional accelerator, or may form a single 256-bit wide port to send or receive 32 byte messages (per cycle) over the NOC via the cluster's Hoplite router. (See Fig. 2.)

Four 2:1 concentrators and a 4x4 crossbar interconnect the eight PEs to the four CRAM ports. In case of simultaneous access to a bank from multiple PEs, an arbiter grants access to one PE and stalls the others.

To send a message, one or more PEs format a 32B message in CRAM, then one PE stores to a memory mapped I/O region. The NOC interface receives the request and atomically sends, in one cycle, a 32 B message to some client of the NOC. If this is another GRVI cluster, the message is received and written into that cluster's CRAM for use by the 8 PEs in that cluster. (Eventually) the same mechanism will be used to store or load 32 B from DRAM or to receive or send a stream of 32 B messages comprising an Ethernet packet to an Ethernet MAC.

If a cluster is configured with one or more accelerators, they can communicate with the PEs via shared memory or directly via a PE's ALU, store-data, and load-data ports.

A cluster may be configured with fewer cores, more or less IRAM and CRAM, to right-size resources for the workload.

The design is optimized for parallel compute kernels (SPMD or MIMD). There is no instruction or data cache. Using multicast Hoplite, all IRAMs in the device can be loaded with a new kernel in 1024 cycles (about 4 µs).

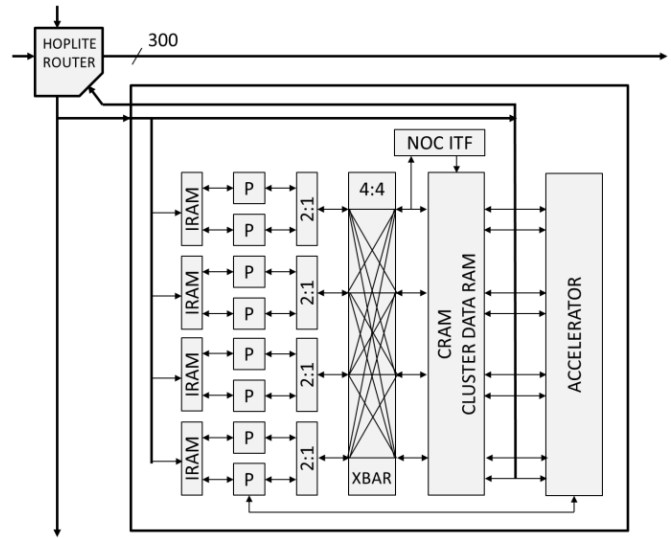

Fig. 2: 8 GRVI Cluster with 288-bit Hoplite interface

### IV. GRVI PHALANX

Fig. 3 is a floorplanned 400-GRVI Phalanx implemented in a Kintex UltraScale KU040. It has 10 rows by 5 columns of clusters (i.e. on a 10x5 Hoplite NOC); each cluster with 8 PEs sharing 32 KB of CRAM. This design uses 73% of the device's LUTs and 100% of its BRAMs. The 300-bit-wide Hoplite NOC uses 6% of the device's LUTs; its amortized cost is less than 40 LUTs/PE. The total size of each processor, its share of the cluster interconnect, and Hoplite router is about 440 LUTs.

In aggregate, the 400 PEs have a peak throughput of about 100,000 MIPS. Total bandwidth into the CRAMs is 600 GB/s. The NOC has a bisection bandwidth of about 700 Gb/s.

Preliminary power data, measured via SYSMON, is about 13 W (33 mW per PE) running a simple test where PE #0 repeatedly receives a request message from every other PE and sends back to each a response message.

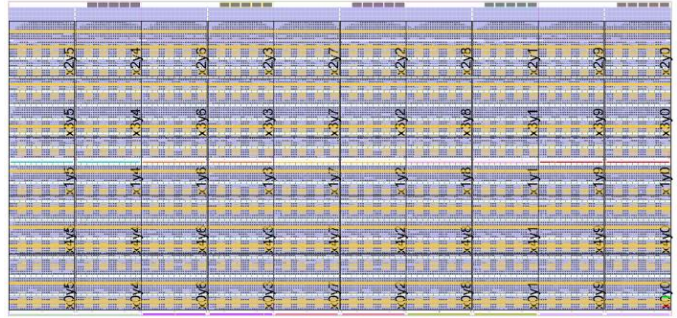

Fig 3: A GRVI Phalanx. 10x5 clusters x 8 PEs = 400 PEs